\documentclass[a4paper]{jpconf}
\usepackage{graphicx}
\def\Up{\uparrow}
\def\Dn{\downarrow}

\def\cI{{\mathcal I}}
\begin{document}
\title{How to distinguish the Haldane/Large-$D$ state and the intermediate-$D$ state
in an $S=2$ quantum spin chain with the $XXZ$ and on-site anisotropies}

\author{Kiyomi~Okamoto${}^1$,
Takashi~Tonegawa$^{2}$,
Hiroki~Nakano$^{3}$,
T\^oru~Sakai$^{3,4,5}$,
Kiyohide~Nomura$^{6}$ and
Makoto~Kaburagi$^{2}$
}

\address{
$^{1}$Department of Physics, Tokyo Institute of Technology, Meguro-ku, Tokyo
152-8551, Japan \\
$^{2}$Professor Emeritus, Kobe University, Kobe 657-8501, Japan\\
$^{3}$Graduate School of Material Science, University of Hyogo, Hyogo
678-1297, Japan \\
$^{4}$Japan Atomic Energy Agency, SPring-8, Hyogo 679-5148, Japan\\
$^{5}$Transformative Research-Project on Iron Pnictide, JST, Saitama
332-0012, Japan \\
$^{6}$Department of Physics, Kyushu University, Fukuoka 812-8581, Japan
}

\ead{kokamoto@phys.titech.ac.jp}

\begin{abstract}
We numerically investigate the ground-state phase diagram of an $S=2$ quantum spin chain
with the $XXZ$ and on-site anisotropies described by
${\mathcal H}=\sum_j (S_j^x S_{j+1}^x+\!S_j^y S_{j+1}^y+\Delta S_j^z S_{j+1}^z) + D \sum_j (S_j^z)^2$,
where $\Delta$ denotes the $XXZ$ anisotropy parameter of the nearest-neighbor
interactions and $D$ the on-site anisotropy parameter. 
We restrict ourselves to the $\Delta>0$ and $D>0$ case for simplicity.
Our main purpose is to obtain the definite conclusion whether there exists or not 
the intermediate-$D$ (ID) phase, which was proposed by Oshikawa in 1992
and has been believed to be absent 
since the DMRG studies in the latter half of 1990's.
In the phase diagram with $\Delta>0$ and $D>0$
there appear the $XY$ state, the Haldane state, the ID state, the large-$D$ (LD) state
and the N\'eel state.
In the analysis of the numerical data
it is important to distinguish three gapped states; the Haldane state, the ID state and the LD state.
We give a physical and intuitive explanation for our level spectroscopy method
how to distinguish these three phases.

\end{abstract}

\section{Introduction}
In this paper, 
using mainly numerical methods, we \cite{tone} investigate the ground-state
phase diagram of the $S=2$ quantum spin chain described by the
Hamiltonian
\begin{equation}
    {\mathcal H}
    = \sum_j (S_j^x S_{j+1}^x + S_j^y S_{j+1}^y + \Delta S_j^z S_{j+1}^z)
            + D \sum_j (S_j^z)^2\,,
    \label{eq:ham}
\end{equation}
where $\Delta$ and $D$ are,
respectively, the $XXZ$ anisotropy parameter of the nn interactions and the
on-site anisotropy parameter.  
We restrict ourselves to the $\Delta>0$ and $D>0$ case for simplicity.

\begin{figure}[h]
\includegraphics[width=12pc]{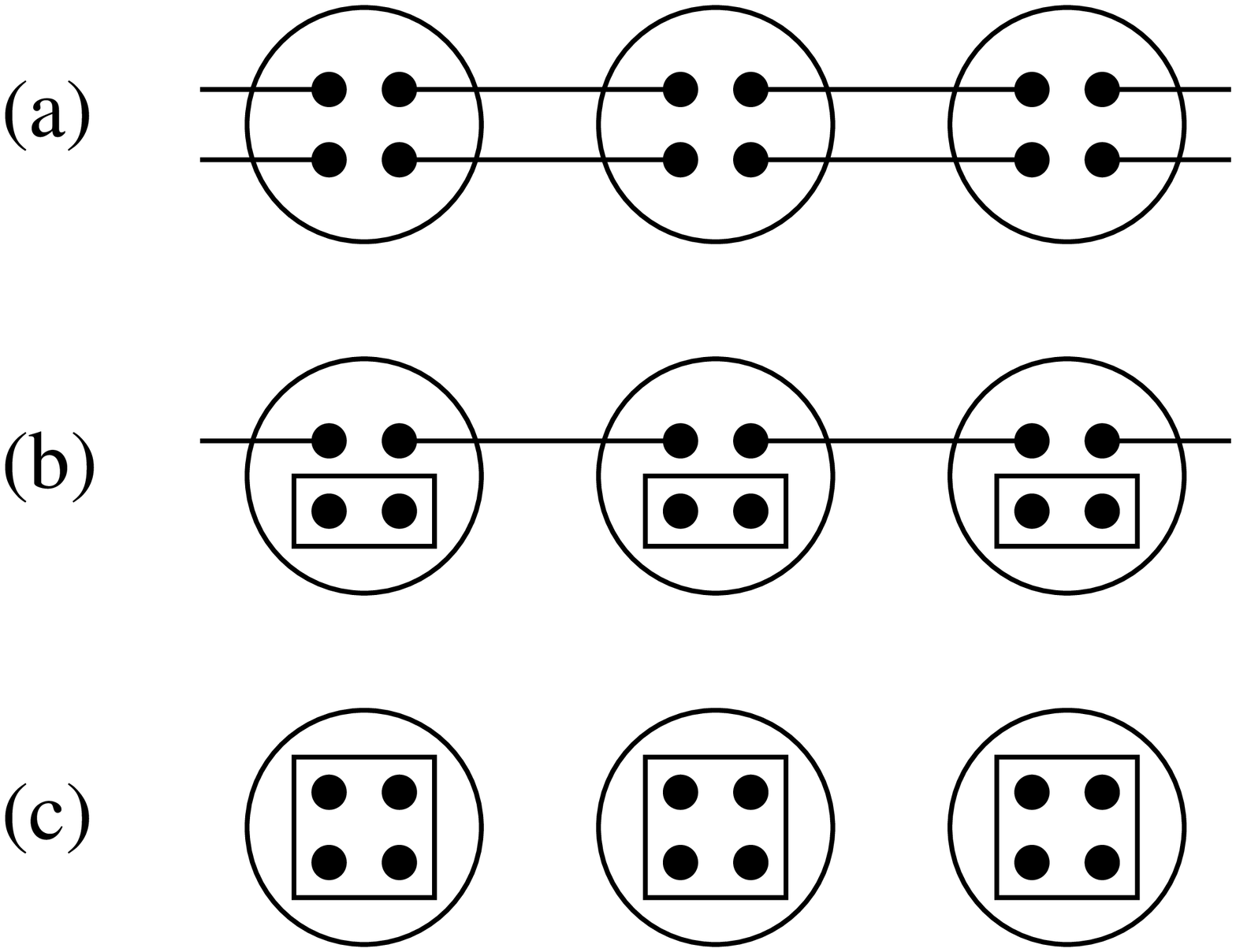}\hspace{2pc}%
\begin{minipage}[b]{18pc}
\caption{\label{fig:vbs-pictures}
Valence bond pictures for (a) the Haldane state, (b) the ID state
   and (c) the LD state.  Big circles denote \hbox{$S\!=\!2$} spins and dots
   \hbox{$S\!=\!1/2$} spins.  Solid lines represent valence bonds
   $\bigl($singlet pairs of two \hbox{$S\!=\!1/2$} spins,
   $(1/\sqrt{2})(\uparrow\downarrow-\downarrow\uparrow)$$\bigr)$.  Two
   \hbox{$S\!=\!1/2$} spins in rectangles are in the
   \hbox{$(S_{\rm tot},S^z_{\rm tot})\!=\!(1,0)$} state and similarly four
   \hbox{$S\!=\!1/2$} spins in squares are in the
   \hbox{$(S_{\rm tot},S^z_{\rm tot})\!=\!(2,0)$} state.}
\end{minipage}
\end{figure}
The ground-state phase diagram of this model was first discussed by Schulz \cite{schulz}.
In his phase diagram for $\Delta>0$ and $D>0$,
there appear the N\'eel phase, the $XY$ phase,
the Haldane phase and the large-$D$ (LD) phase. 
In 1992 Oshikawa \cite{oshikawa92}
predicted, for $S \ge 2$ integer quantum spin cases, 
the existence of the intermediate-$D$ (ID) phase, 
the valence bond picture of which is depicted
in Fig.\ref{fig:vbs-pictures}(b).  
After that, however,
by use of the density-matrix renormalization-group (DMRG)
calculation, Schollw\"ock et al. \cite{schollwoeck1,schollwoeck2} and Aschauer
and Schollw\"ock \cite{aschauer} concluded the absence of the ID phase.
By use of the level spectroscopy (LS)
analysis of numerical results of exact-diagonalization calculations, 
Nomura and Kitazawa \cite{nomura} showed in the case of $\Delta=1$ that, with the
increase of $D$ from zero, the ground state changes
as Haldane state $\Rightarrow$ $XY$ state $\Rightarrow$ LD state.
The first and second transitions occur at $D_{\rm c1}=0.043$ and
$D_{\rm c2}=2.39$, respectively.
Since these works
it has been believed for a long time
that the ID phase does not exist in the phase diagram of the present $S =2$ model.  

The DMRG is a very powerful method for spin chains,
especially when the magnitude of the spin gap is targeted.
However, it is difficult to deal with the phase transition in some cases, 
because the phase transition point is determined by the extrapolated
values of the finite-size being equal to zero.
Thus, it is somewhat hard to obtain accurate results
if the magnitude of the gap is very small
(i.e. the correlation length is very long).
This difficulty of zero-or-finite problem is not special to the DMRG method
but is common to almost all numerical methods.
On the other hand, the LS method is conspicuous on this point
because the phase transition point is determined from the
crossing point of two related excitations,
which is free from the zero-or-finite problem.

The LS method \cite{nomura,kitazawa} is firmly based on the effective Hamiltonian, 
renormalization group method and conformal field theory.
In this paper 
we give a physical and intuitive explanation for our LS method
how to distinguish the Haldane, ID and LD states
sketched in Fig.\ref{fig:vbs-pictures}.

\section{A very simple example: $S=1/2$ chain with bond alternation}

\begin{figure}[h]
\begin{minipage}{18pc}
  \scalebox{0.3}{\includegraphics{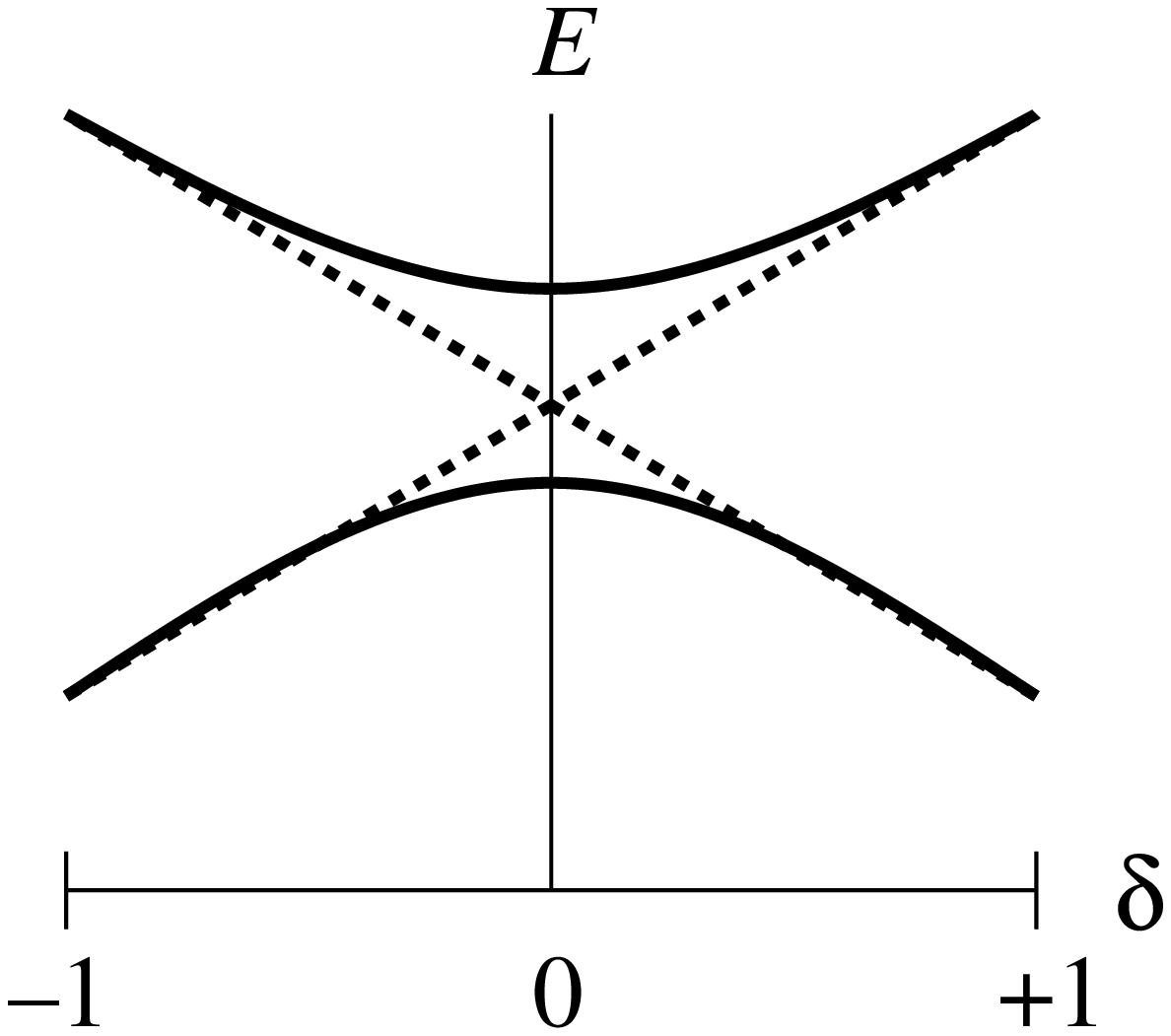}}
  \caption{\label{fig:twist-2}
  Schematic behaviors of $\psi_0^{\rm PBC}(\delta>0)$ and $\psi_0^{\rm PBC}(\delta<0)$ 
  under the PBC.
  They do not cross with each other (not broken lines).
  They smoothly changes with each other
  because the mixing of these two states occurs.}
\end{minipage}\hspace{2pc}%
\begin{minipage}{18pc}
  \scalebox{0.3}{\includegraphics{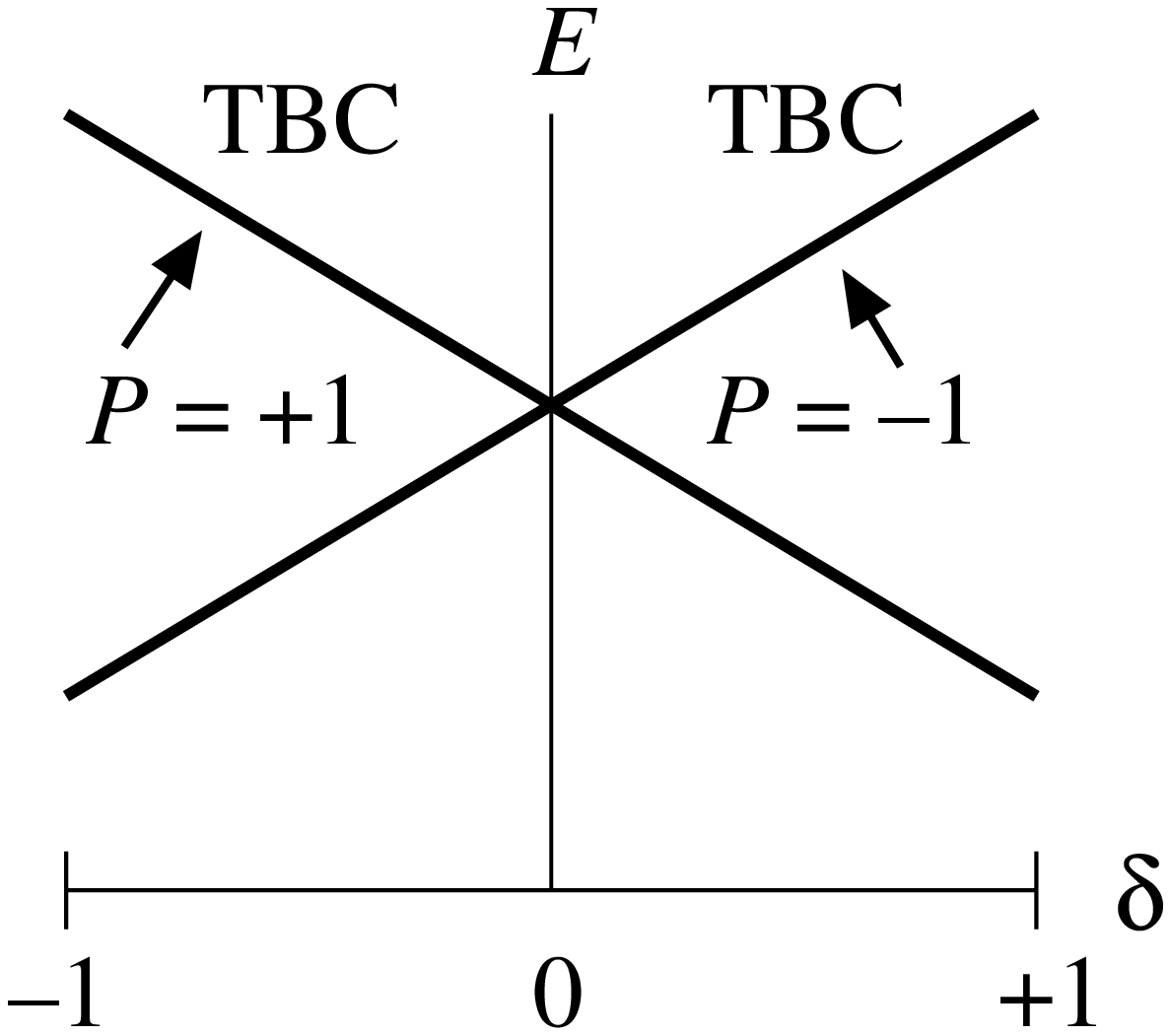}}
  \caption{\label{fig:twist-4}
  Schematic behaviors of $\psi_0^{\rm PBC}(\delta>0)$ and $\psi_0^{\rm PBC}(\delta<0)$ 
  under the TBC.
  They cross with each other at the transition point $\delta=0$.
  The mixing of these two states does not occur because they are protected 
  by different $P$.}
\end{minipage} 
\end{figure}

Before discussing $S=2$ chain problem,
let us visit a very simple example of an $S=1/2$ chain with bond alternation 
described by
\begin{equation}
    {\mathcal H}_{\rm ba}(\delta)
    = \sum_j [1 + (-1)^{j+1}\delta ]{\bi S}_j  \cdot {\bi S}_{j+1},~~~~~S=1/2
    \label{eq:bond-alt}
\end{equation}
where $\delta$ is the bond-alternation parameter ($-1 \le \delta \le 1$).
For $\delta=0$ the ground state of the Hamiltonian (\ref{eq:bond-alt}) is the Tomonaga-Luttinger liquid
state.
We treat an $N=8$ system for a while.
When $\delta=\pm 1$ the ground states of the Hamiltonian (\ref{eq:bond-alt}) 
under the periodic boundary condition (PBC) are trivial
\begin{eqnarray}
    &&\psi_0^{\rm PBC}(\delta=+1) = [1,2]\,[3,4] \,[5,6]\, [7,8] \\
    &&\psi_0^{\rm PBC}(\delta=-1) = [2,3]\,[4,5] \,[6,7]\,[8,1] 
\end{eqnarray}
where $[i,j] \equiv (1/\sqrt{2})(\Up_i \Dn_j - \Dn_i \Up_j) = - [j,i]$.
It is well known that the ground state for $\delta>0$, $\psi_0^{\rm PBC}(\delta>0)$, 
is similar to $\psi_0^{\rm PBC}(\delta=+1)$
and $\psi_0^{\rm PBC}(\delta<0)$ is similar to $\psi_0^{\rm PBC}(\delta=-1)$.
The quantum phase transition occurs at $\delta=0$.
The state $\psi_0^{\rm PBC}(\delta=\pm1)$ is the highest energy states at $\delta=\mp1$, respectively.
However, since the crossing of the ground state energy does not occur when $\delta$ is swept,
as depicted in Fig.\ref{fig:twist-2},
we cannot determine the transition point from the crossing.
Because $\psi_0^{\rm PBC}(\delta=\pm1)$ are not distinguished by the eigenvalue
of the discrete symmetry,
there occurs the mixing of these two states resulting in the level repulsion.
In other words, the degeneracy is lifted by the {\lq\lq}perturbation".

Let us impose the twisted boundary condition (TBC)
\begin{equation}
    {\bi S}_8  \cdot {\bi S}_{1}
    \Rightarrow
    -S_8^x S_1^x - S_8^y S_1^y + S_8^z S_1^z.
\end{equation}
In this case
the ground state of Hamiltonian (\ref{eq:bond-alt}) for $\delta=\pm 1$ cases are, respectively
\begin{eqnarray}
    &&\psi_0^{\rm TBC}(\delta=+1) = [1,2]\,[3,4] \,[5,6]\, [7,8] \\
    &&\psi_0^{\rm TBC}(\delta=-1) = [2,3]\,[4,5] \,[6,7]\,\{8,1\} 
\end{eqnarray}
where $\{i,i\} \equiv (1/\sqrt{2})(\Up_i \Dn_j + \Dn_j \Up_i) = \{j,i\}$.
Here we introduce the lattice inversion operator ${\cal I}$ about the bond center of
the boundary,
which works $1 \leftrightarrow 8,\,2 \leftrightarrow 7,\,3 \leftrightarrow 6,\,4 \leftrightarrow 5$.
From
\begin{eqnarray}
    {\cal I} \psi_0^{\rm TBC}(\delta=+1)
     &=& [8,7]\,[6,5] \, [4,3]\,[2,1] 
     = + \psi_0^{\rm TBC}(\delta=+1) \\
    {\cal I} \psi_0^{\rm TBC}(\delta=-1)
     &=& [7,6]\,[5,4] \,[3,2]\,\{1,8\} 
     = - \psi_0^{\rm TBC}(\delta=-1) 
\end{eqnarray}
we see that $\psi_0^{\rm TBC}(\delta=\pm 1)$ have different parity eigenvalues $P = \pm 1$.
Then the states $\psi_0^{\rm TBC}(\delta >0)$ and $\psi_0^{\rm TBC}(\delta < 0)$
have different parities $P$
since they are adiabatically connected to
$\psi_0^{\rm TBC}(\delta=+1)$ and $\psi_0^{\rm TBC}(\delta=+1)$, respectively. 
The Hamiltonian (\ref{eq:bond-alt}) is $P$-invariant,
which means there is no mixing between the states with different $P$.
Thus the lowest $P=+1$ state and $P=-1$ state crosses with each other
at the quantum phase transition point $\delta=0$,
as demonstrated in Fig.\ref{fig:twist-4}.
We note that $\psi_0(\delta >0)$ and $\psi_0(\delta <0)$ have same
eigenvalues $P$ for the PBC case.

\section{$S=2$ case}

Here we explain the $S=2$ case,
considering $N=4$ system for simplicity.
Under the PBC the Haldane state and the ID state corresponding to the valence bond pictures
are, respectively
\begin{eqnarray}
    &&\psi_{\rm H}^{\rm PBC}
    = [1,2]^2\,[2,3]^2\,[3,4]^2\,[4,1]^2 \\
    &&\psi_{\rm ID}^{\rm PBC}
    = [1,2]\,[2,3]\,[3,4]\,[4,1]
\end{eqnarray}
where two valence bonds between the $i$th and $j$th spins are
abbreviated as $[i,j]^2$.
Although these wave functions are not exact ones in general,
the exact wave functions for the Haldane and ID states
are adiabatically connected to above two wave functions respectively.
When we operate $\cI$ on these states,
we obtain
\begin{eqnarray}
    &&\cI\psi_{\rm H}^{\rm PBC}
    = [4,3]^2\,[3,2]^2\,[2,1]^2\,[1,4]^2 
    = \psi_{\rm H}^{\rm PBC}  \\
    &&\cI\psi_{\rm ID}^{\rm PBC}
    = [4,3]\,[3,2]\,[2,1]\,[1,4]
    = \psi_{\rm ID}^{\rm PBC}
\end{eqnarray}
Then, $P=+1$ for both the Haldane state and the ID state.
Since there are no valence bonds in the LD state,
it is clear $P=+1$ for the LD state.
We cannot distinguish these three states under the PBC.
\begin{figure}[h]
\begin{minipage}{17pc}
  \scalebox{0.3}{\includegraphics{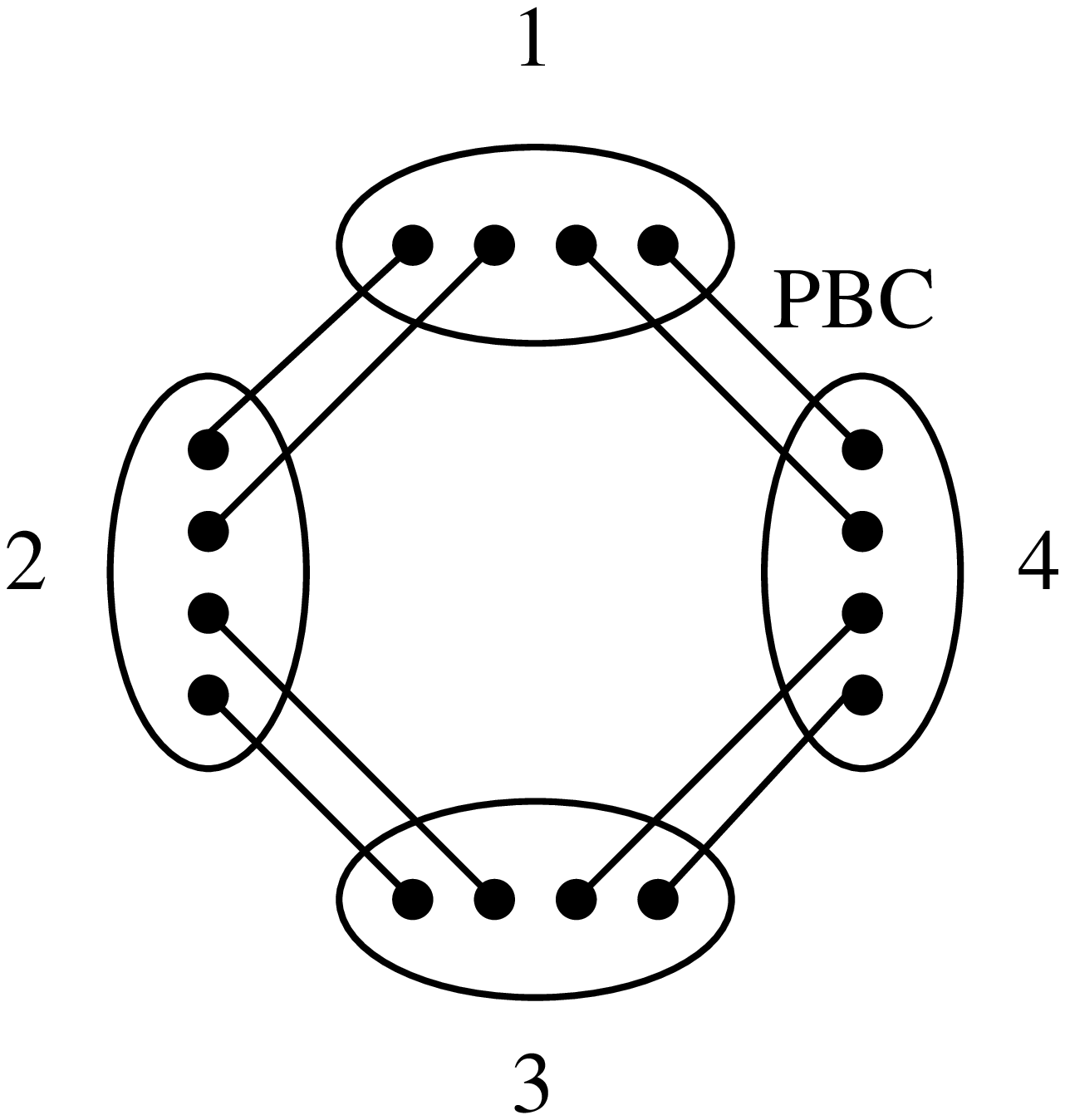}}
  \caption{\label{fig:s=2-haldane-pbc}
  Haldane state under the PBC.}
\end{minipage}\hspace{2pc}%
\begin{minipage}{17pc}
  \scalebox{0.3}{\includegraphics{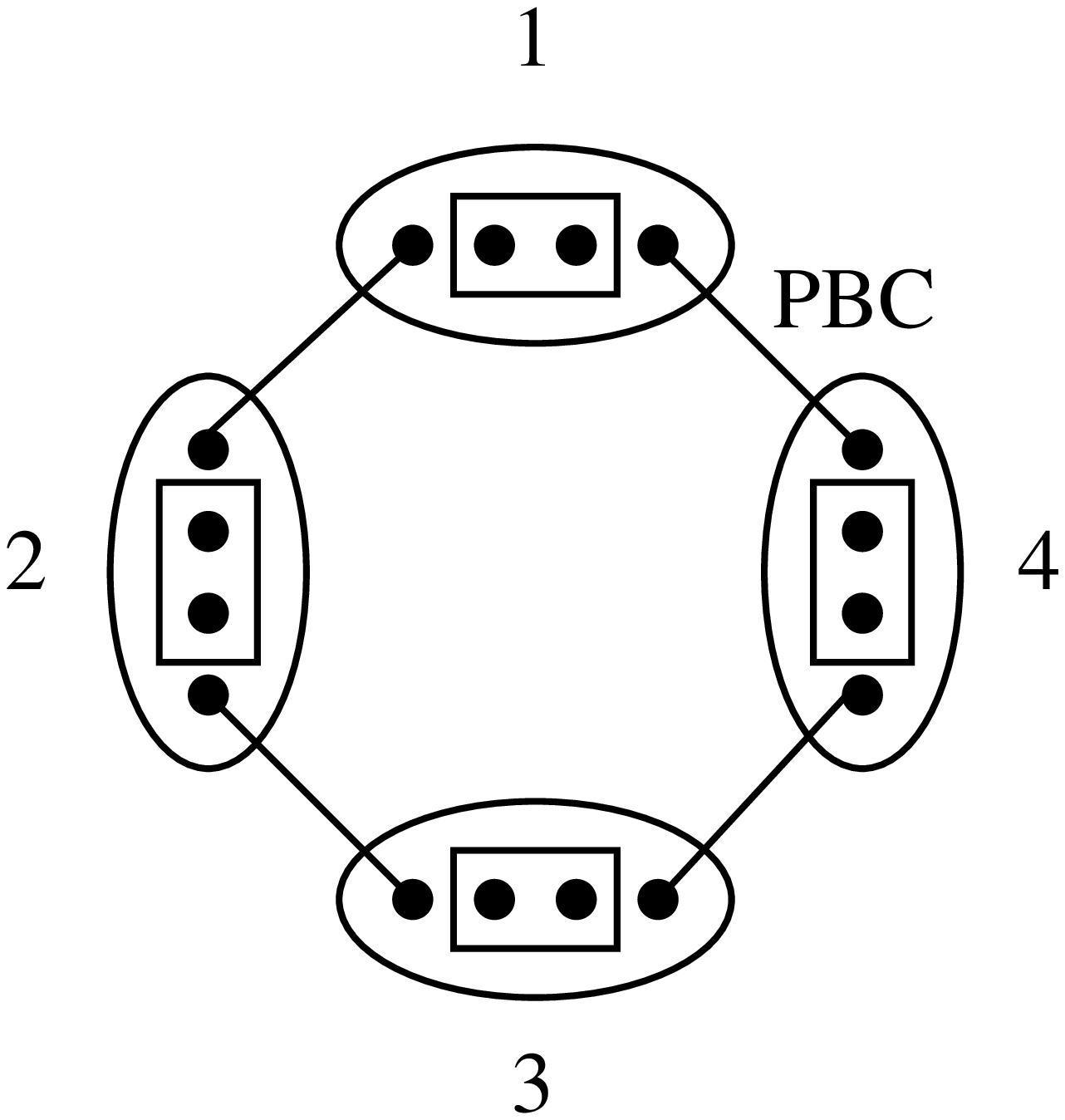}}
  \caption{\label{fig:s=2-id-pbc}ID state under the PBC.}
\end{minipage} 
\end{figure}

Under the TBC the Haldane state and the ID state corresponding to the valence bond picture
are, respectively
\begin{eqnarray}
    &&\psi_{\rm H}^{\rm TBC}
    = [1,2]^2\,[2,3]^2\,[3,4]^2\,\{4,1\}^2 \\
    &&\psi_{\rm ID}^{\rm TBC}
    = [1,2]\,[2,3]\,[3,4]\,\{4,1\}
\end{eqnarray}
When we operate $\cI$ on these states,
we obtain
\begin{eqnarray}
    &&\cI\psi_{\rm H}^{\rm TBC}
    = [4,3]^2\,[3,2]^2\,[2,1]^2\,\{1,4\}^2 
    = \psi_{\rm H}^{\rm TBC}  \\
    &&\cI\psi_{\rm ID}^{\rm TBC}
    = [4,3]\,[3,2]\,[2,1]\,\{1,4\}
    = -\psi_{\rm ID}^{\rm TBC}
\end{eqnarray}
Then, $P=+1$ for the Haldane state and $P=-1$ for the ID state.
Since there are no valence bonds in the LD state,
it is clear $P=+1$ for the LD state.
Thus we can distinguish the ID state from the Haldane and LD states
by the eigenvalue $P$.
Namely, under the TBC, 
if the lowest eigenstate is $P=-1$, the ground state is the ID state.
On the other hand, 
if the lowest eigenstate is $P=+1$, the ground state is the Haldane state or the LD state.
\begin{figure}[h]
\begin{minipage}{17pc}
  \scalebox{0.3}{\includegraphics{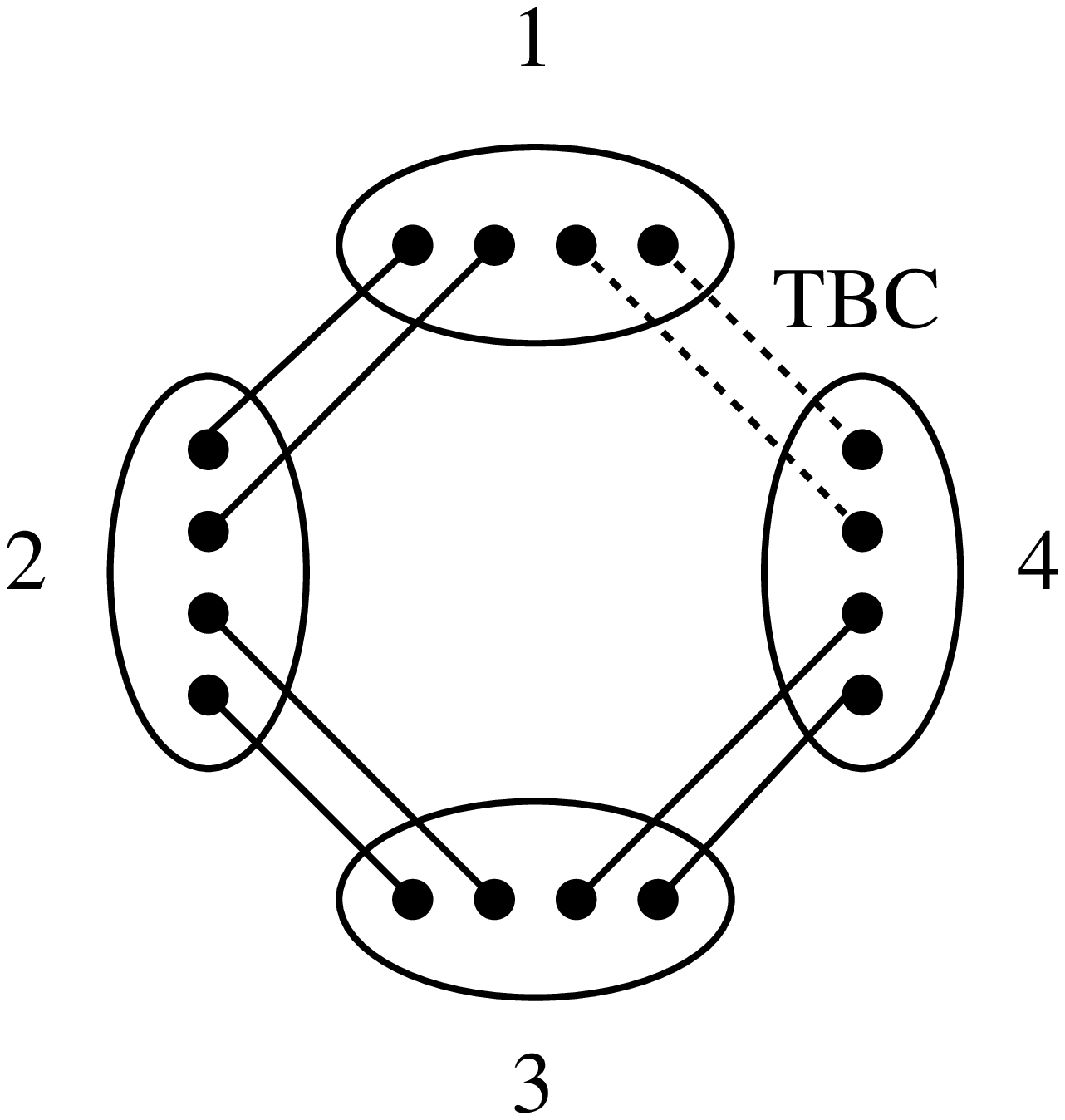}}
  \caption{\label{fig:s=2-haldane-tbc}
  Haldane state under the TBC.
  Broken lines denote $\{i,j\}$.}
\end{minipage}\hspace{2pc}%
\begin{minipage}{17pc}
  \scalebox{0.3}{\includegraphics{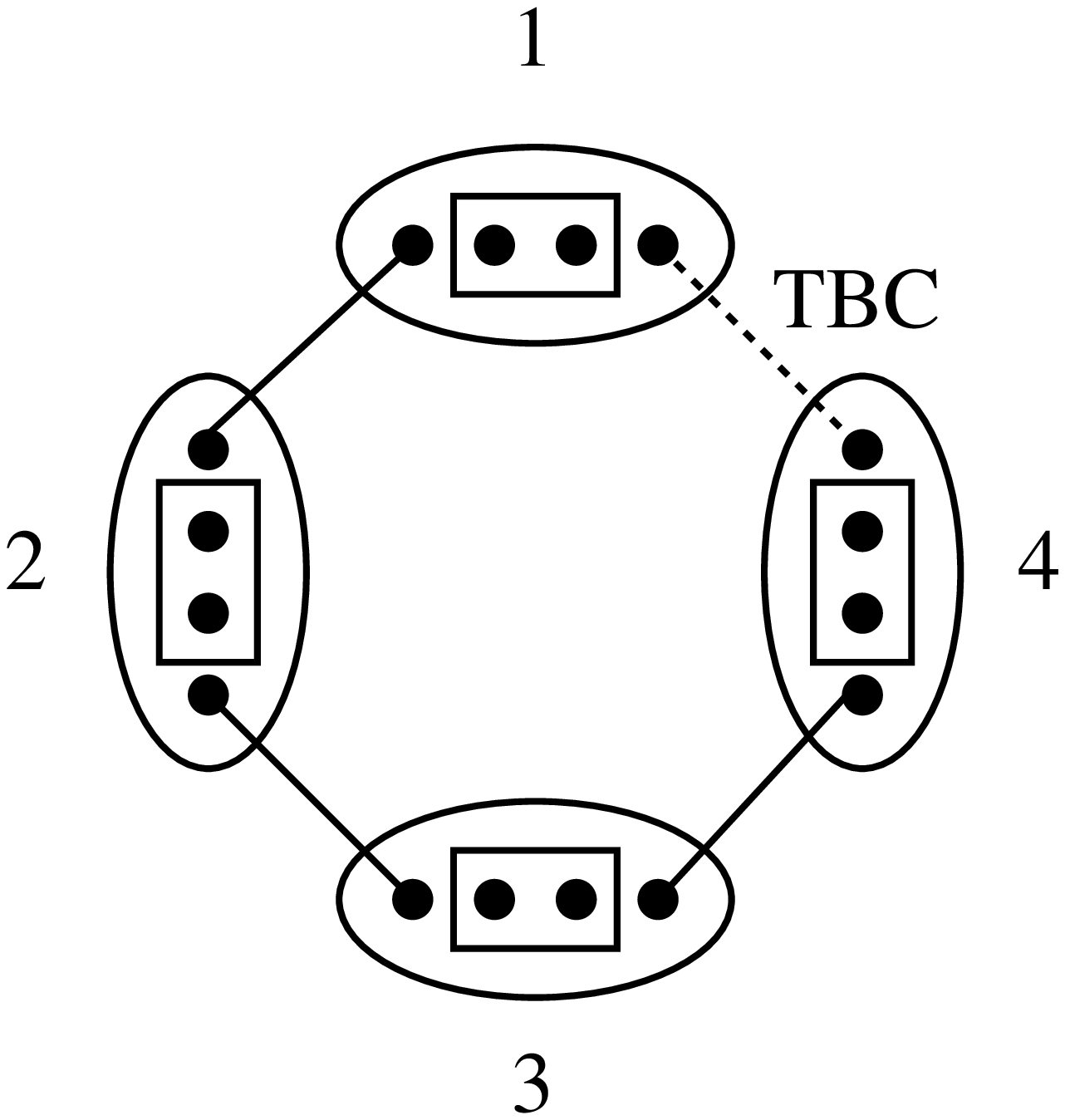}}
  \caption{ID state under the TBC.
  Broken line denotes $\{i,j\}$.}
\end{minipage} 
\end{figure}

There may arise a question how to distinguish the Haldane state and the LD state.
Recently Pollmann, Berg, Turner and Oshikawa \cite{pollmann1}
stated that the Haldane state is essentially indistinguishable from the LD state.
They also constructed a one-parameter matrix product state
which interpolates the Haldane state and the LD state without any quantum phase
transition.
Our results \cite{tone} also indicate that the Haldane state and the LD state
belong to the same phase, as shown later in Figs. 10 and 11.

\section{Ground state phase diagram of an $S=2$ chain}

From the discussion of the previous section,
we should compare the two energy levels
$E_0^{\rm TBC}(M=0,P=+1)$ and $E_0^{\rm TBC}(M=0,P=-1)$,
where $E_0^{\rm TBC}(M,P)$ is the lowest energy in the subspace
of magnetization $M = \sum_j S_j^z$ and parity $P$ under the TBC.
To obtain the phase diagram,
we further check whether the ground state is gapless ($XY$) state or gapped
state (Haldane, ID or LD).
Nomura and Kitazawa \cite{nomura} showed that
the ground state is gapless when $E_0^{\rm PBC}(M=2) < E_0^{\rm TBC}(M=0,P)$,
where $E_0^{\rm PBC}(M=2)$ is the lowest energy in the subspace
$M = 2$ under the PBC.
This condition was obtained through the effective Hamiltonian 
and the renormalization group method,
unfortunately for which we have no physical and intuitive explanation.
Anyway, we have to compare three energy levels,
$E_0^{\rm TBC}(M=0,P=+1)$, $E_0^{\rm TBC}(M=0,P=-1)$ and $E_0^{\rm PBC}(M=2)$.
Namely, the ground state is the Haldane/LD (H/LD) state, the ID state or the $XY$ state
according as $E_0^{\rm TBC}(M=0,P=+1)$, 
$E_0^{\rm TBC}(M=0,P=-1)$ or 
$E_0^{\rm PBC}(M=2)$ is the lowest.
\begin{figure}[h]
\begin{minipage}{18pc}
  \scalebox{0.22}{\includegraphics{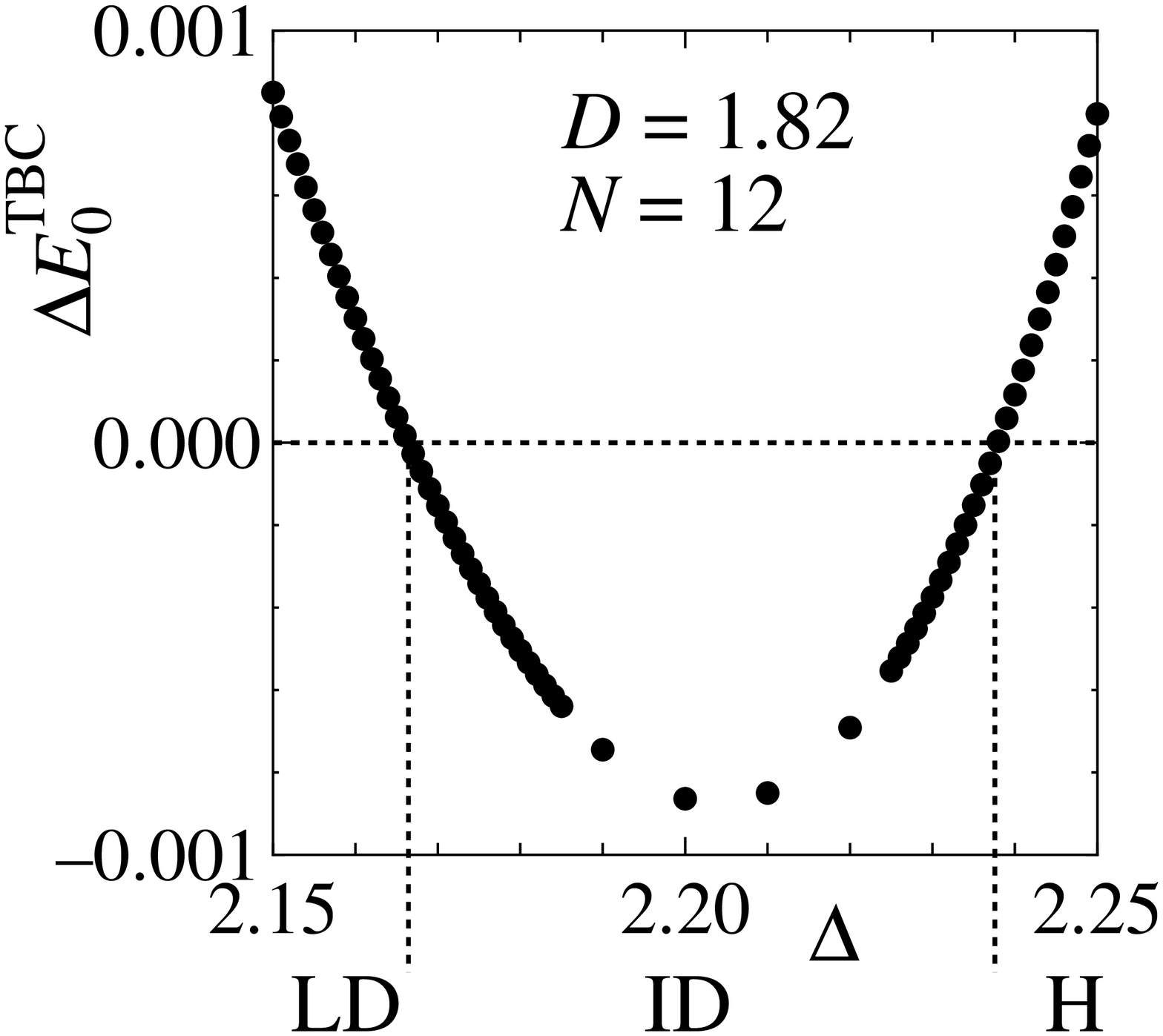}}
  \caption{\label{fig:h-id-n=12}
  Behavior of $\Delta E_0^{\rm TBC}\! \equiv \! E_0^{\rm TBC}(M\!=\!0,P\!=\!+1)\! -\! E_0^{\rm TBC}(M\!=\!0,P\!=\!-1)$.}
\end{minipage}\hspace{2pc}%
\begin{minipage}{18pc}
  \scalebox{0.22}{\includegraphics{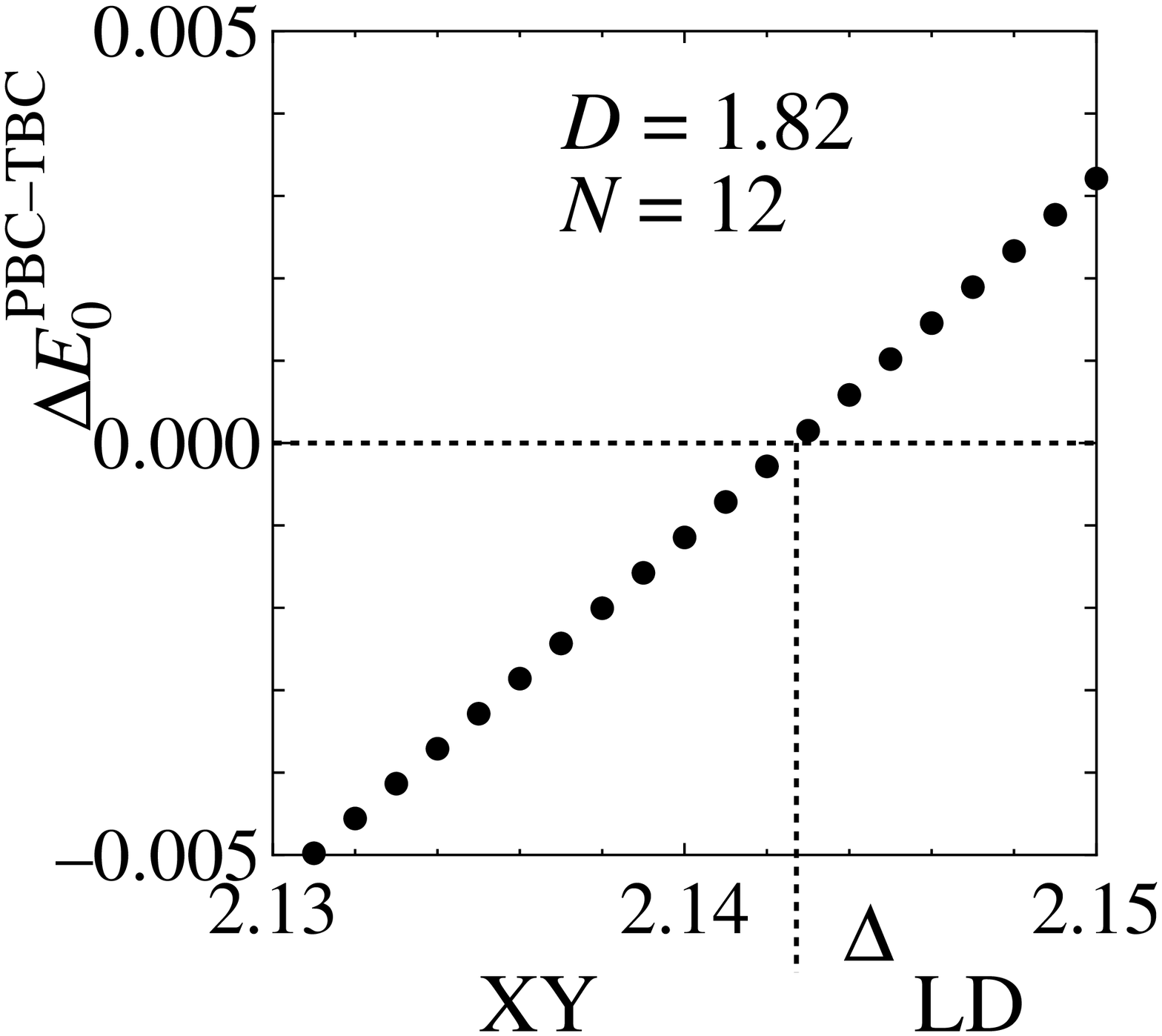}}
  \label{fig:h-xy-n=12}
  \caption{Behavior of $\Delta E_0^{\rm PBC-TBC} \!
          \equiv\! E_0^{\rm PBC}(M\!=\!2)\! -\! E_0^{\rm TBC}(M\!=\!0,P\!=\!+1)$.}
\end{minipage} 
\end{figure}

Figure \ref{fig:h-id-n=12} shows
$\Delta E_0^{\rm TBC} \equiv E_0^{\rm TBC}(M=0,P=+1) - E_0^{\rm TBC}(M=0,P=-1)$
as a function of $\Delta$ when $D=1.82$ obtained by the numerical diagonalization
of 12 spin systems.
From this figure we can determine the phase boundary
between the Haldane/LD phase and the ID phase.
We note that $E_0^{\rm PBC}(M=2)$ is higher than $E_0^{\rm TBC}(M=0,P=\pm1)$
in this region.
The behavior of $\Delta E_0^{\rm PBC-TBC} \equiv E_0^{\rm PBC}(M=2) - E_0^{\rm TBC}(M=0,P=+1)$
is shown in Fig. 9,
from which the phase boundary between the $XY$ phase and Haldane/LD phases is determined.
The energy $E_0^{\rm TBC}(M=0,P=-1)$ is higher in this region.
Examples of the size dependence of the critical values of $\Delta$ are summarized in Table 1, from which we see that the finite-size effects are not so serious
in our LS anayses.
Since the transition between the Haldane/LD phase and the N\`eel phase is expected to of the Ising type,
we use the phenomenological renormalization group method for the numerical analysis \cite{tone}.
\begin{table}[h]
\caption{Examples of critical values of $\Delta$ in the case of $D=1.82$ obtained by the LS method.}
\label{table}
\begin{center}
\begin{tabular}{|r|l|l|l|}
   \hline
     N  &$\Delta_{\rm c}^{({\rm H,ID})}(N)$ &$\Delta_{\rm c}^{({\rm ID,LD})}(N)$ &$\Delta_{\rm c}^{(XY,{\rm LD})}(N)$ \\ \hline    
     6  &2.17687                           &2.14971                            &2.08931  \\ \hline   
     8  &2.22262                           &2.16106                            &2.12246  \\ \hline
    10  &2.23529                           &2.16527                            &2.13607  \\ \hline
    12  &2.23793                           &2.16639                            &2.14265  \\ \hline
    $\infty$ &$2.241 \pm 0.001$             &$2.167 \pm 0.001$                   &$2.156 \pm 0.001$ \\ \hline
\end{tabular}
\end{center}
\end{table}
\begin{figure}[h]
\begin{minipage}{17pc}
  \scalebox{0.3}{\includegraphics{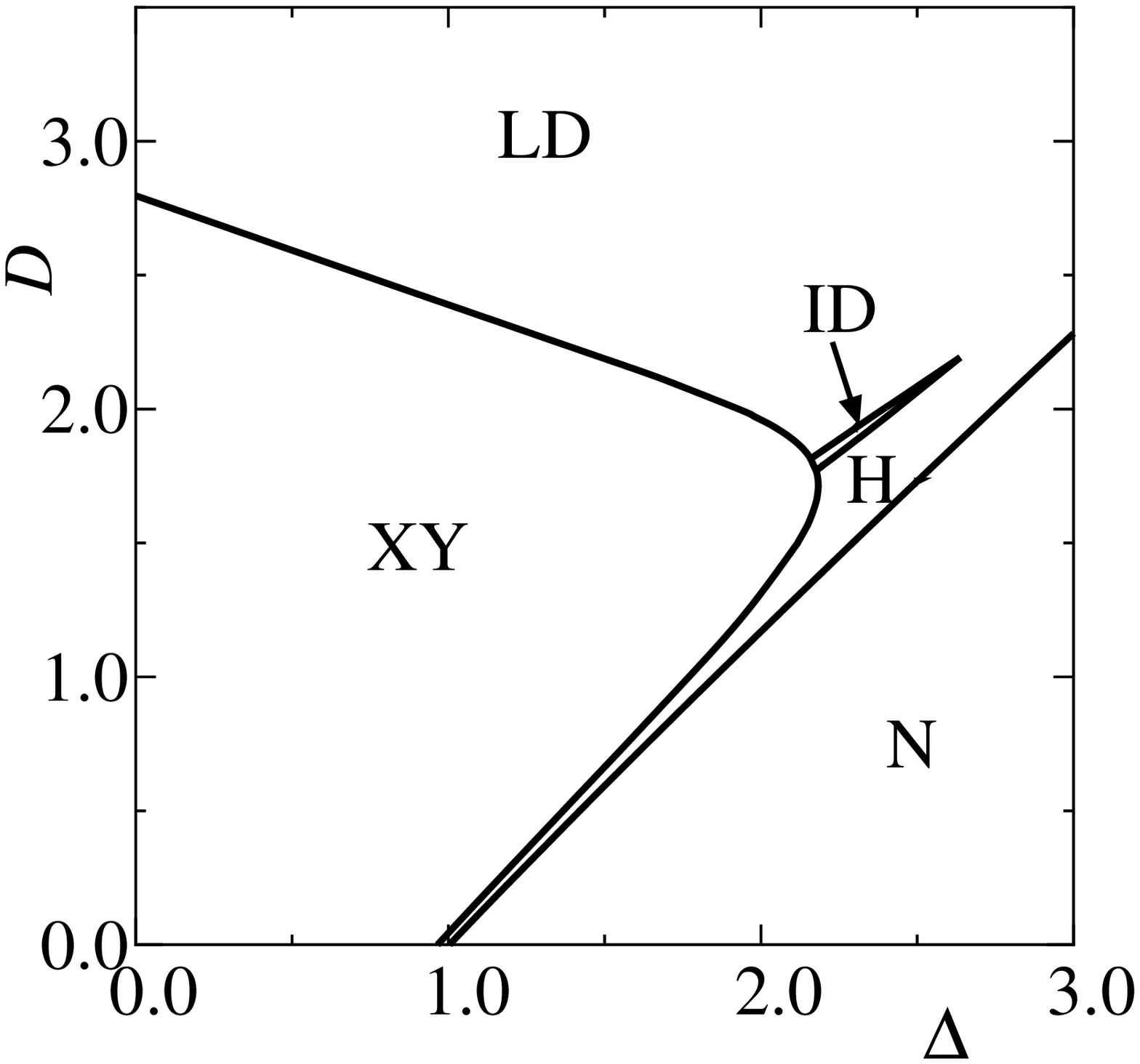}}
  \caption{\label{fig:fig4a-new}
  Phase diagram of an $S=2$ chain.
  H and N are the abbreviations of Haldane and N\'eel, respectively.}
\end{minipage}\hspace{2pc}%
\begin{minipage}{16pc}
  \scalebox{0.3}{\includegraphics{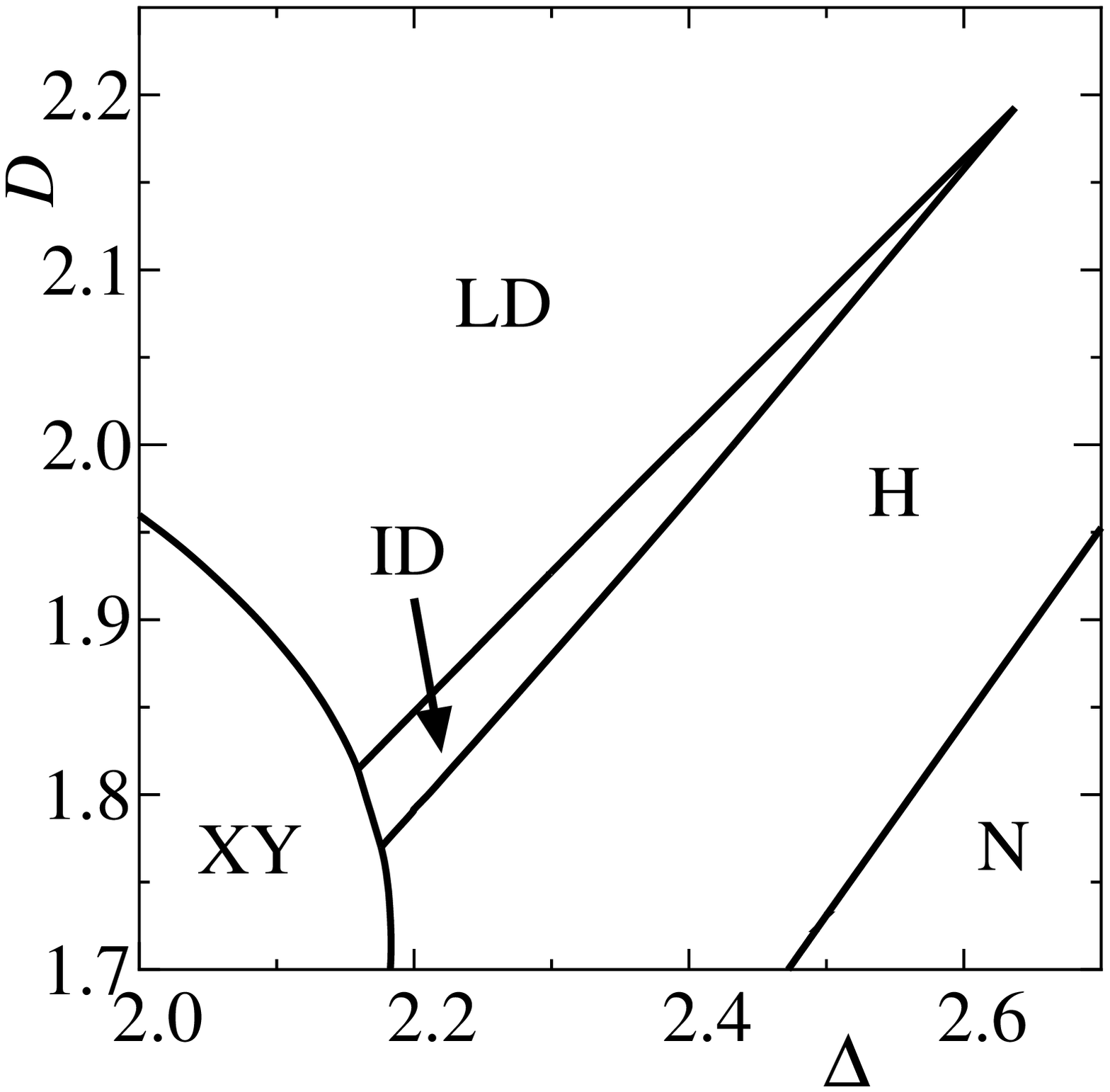}}
  \label{fig:fig4b-new}
  \caption{Enlarged phase diagram near the ID phase. \\}
\end{minipage} 
\end{figure}

Our final phase diagram is shown in Figs. \ref{fig:fig4a-new} and 11.
The remarkable natures of the phase diagram are:~(1)
there exists the ID phase which was predicted by Oshikawa in
    1992 and has been believed to be absent for a long time;
(2) the Haldane state and the LD state belong to the same phase.

\ack

We would like to express our appreciation to Professor Masaki Oshikawa 
and Dr. Frank Pollmann for their invaluable discussions and comments.
We also thank the Supercomputer Center, Institute for Solid
State Physics, University of Tokyo, the Computer Room, Yukawa Institute for
Theoretical Physics, Kyoto University and the Cyberscience Center,
Tohoku University for computational facilities.  The present work has
been supported in part by a Grant-in-Aid for Scientific Research (B)
(No.~20340096), and a Grant-in-Aid for Scientific Researches on Priority Areas
^^ ^^ Novel states of matter
induced by frustration'' from the Ministry of Education, Culture, Sports,
Science and Technology.

\end{document}